\documentclass{aa}
\usepackage{graphicx}
\usepackage{txfonts}
\usepackage[authoryear]{natbib}
\bibpunct{(}{)}{;}{a}{}{,}

\begin{document}

\title{The changing milliarcsecond radio morphology \\ of the gamma-ray binary LS~5039}


\author{M. Rib\'o\inst{1}, J.M. Paredes\inst{1}, J. Mold\'on\inst{1}, J. Mart\'\i \inst{2}, M. Massi\inst{3}}
\institute{Departament d'Astronomia i Meteorologia, Universitat de Barcelona, 
Mart\'{\i} i Franqu\`es 1, 08028 Barcelona, Spain\\
\email{mribo@am.ub.es; jmparedes@ub.edu; jmoldon@am.ub.es}
\and Departamento de F\'{\i}sica (EPS), Universidad de Ja\'en,
Campus Las Lagunillas s/n, 23071 Ja\'en, Spain
\and Max Planck Institut f\"ur Radioastronomie, Auf dem H\"ugel 69, 53121 Bonn, Germany
}

\authorrunning{Rib\'o et~al.}

\date{Received / Accepted}

\abstract
{LS~5039 is one of the few TeV emitting X-ray binaries detected so
far. The powering source of its multiwavelength emission can be accretion in a
microquasar scenario or wind interaction in a young nonaccreting pulsar
scenario.}
{To present new high-resolution radio images and compare them with the
expected behavior in the different scenarios.}
{We analyze Very Long Baseline Array (VLBA) radio observations that provide
morphological and astrometric information at milliarcsecond scales.}
{We detect a changing morphology between two images obtained five days apart. 
In both runs there is a core component with a constant flux density, and
an elongated emission with a position angle (PA) that changes by $12\pm3\degr$
between both runs. The source is nearly symmetric in the first run and
asymmetric in the second one. The astrometric results are not conclusive.}
{A simple and shockless microquasar scenario cannot easily explain the observed
changes in morphology. An interpretation within the young nonaccreting pulsar
scenario requires the inclination of the binary system to be very close to the
upper limit imposed by the absence of X-ray eclipses.}


\maketitle

\section{Introduction} \label{introduction}

Very high energy (VHE) gamma-ray emission in the TeV range has been detected in
four massive X-ray binaries. PSR~1259$-$63 contains a young
nonaccreting millisecond (ms) radio pulsar \citep{aharonian05a}. The
accreting/ejecting microquasar Cygnus~X-1 has recently been found to
be a TeV emitter \citep{albert07}. LS~I~+61~303 \citep{albert06},
suggested to be a fast precessing microquasar \citep{massi04}, displays a
changing milliarcsecond radio morphology that \cite{dhawan06} interpreted in
the context of the interaction between the wind of the companion and the
relativistic wind of a young nonaccreting ms pulsar \citep{dubus06}. Finally,
the nature of the powering source in LS~5039, whether accretion or
rotation, is unknown (see \citealt{aharonian05b},\citeyear{aharonian06} for the
TeV results).

The system LS~5039 contains a compact object of unknown nature, with
mass between 1.4 and 5 M$_\odot$, orbiting every 3.9~days an ON6.5\,V((f))
donor \citep{casares05}. The detection of elongated asymmetric emission in
high-resolution radio images obtained with the Very Long Baseline Array (VLBA)
and the European VLBI Network (EVN) was interpreted as evidence of its
microquasar nature, and suggested that the source was persistently producing
mildly relativistic ejections with a velocity of $\sim$0.15$c$
\citep{paredes00,paredes02}. Although the X-ray spectra are compatible with
those of accreting black holes during the so-called low/hard state
\citep{bosch05}, the radio spectra are optically thin with a spectral index of
$-$0.5 \citep{marti98,ribo99}. Theoretical modelling in the microquasar
scenario has allowed researchers to reproduce the observed spectral energy
distribution (SED) from radio to VHE gamma-rays \citep{paredes06}. However, the
lack of clear accretion signatures and the similarities with the SEDs of
PSR~1259$-$63 and LS~I~+61~303 has led other authors to model
its multiwavelength emission using the scenario of wind interactions
\citep{dubus06}. One of the predictions of this kind of modelling is the
periodic change in the direction and shape of the extended radio morphology as
well as in the peak position of the radio core, depending on the orbital phase.

In this paper, we report on new high-resolution radio images of
LS~5039 in an attempt to disentangle the two possible scenarios using
morphological and astrometric information.

\section{Observations} \label{observations}

We observed LS~5039 with the National Radio Astronomy Observatory
(NRAO) VLBA and the Very Large Array (VLA) at 5~GHz frequency on 2000 June 3
and 8. We used the VLA in its C configuration, both as a connected
interferometer and as a phased array. The two observing sessions, hereafter
run~A and run~B, spanned from 4:30 to 12:30~UT on the corresponding dates, and
were thus centered on MJD~51698.4 and MJD~51703.4, respectively. The orbital
phases of the system were in the range 0.43--0.51 for run~A and in the range
0.71--0.79 of the following orbital cycle for run~B (using the ephemeris from
\citealt{casares05}).

We performed the observations using the phase-referencing technique, switching
between the phase reference calibrator J1825$-$1718 and LS~5039, separated
2\fdg47, with cycling times of 5.5 minutes, compatible with the expected
coherence time. The fringe finder was 3C~345. The ICRF source J1911$-$2006,
located at 11\fdg9 from LS~5039, was observed every 22 minutes to monitor the
performance of the observations. To check the stability of the astrometry we
observed the source J1837$-$1532, located at 3\fdg38 from the phase reference
calibrator J1825$-$1718.

We recorded the data with 2-bit sampling at 256~Mbps, at left-hand circular
polarization. A total bandwidth of 64~MHz was provided by 8 sub-bands. The data
were processed at the VLBA correlator in Socorro, using an integration time of
4~s.

The position used hereafter for the phase reference source
J1825$-$1718, obtained by means of dedicated geodetic Very Long
Baseline Interferometry (VLBI) observations, is $\alpha_{\rm J2000.0}=18^{\rm
h} 25^{\rm m} 36\fs53237\pm0\fs00021$ (or $\pm3.0$~mas) and $\delta_{\rm
J2000.0}=-17\degr 18\arcmin 49\farcs8534\pm0\farcs0045$ (or $\pm4.5$~mas) in
the frame of ICRF-Ext.1 (Craig Walker, private communication). However, this
information was not available at the time of correlation, which was performed
for a calibrator position shifted by $\Delta\alpha=+57.0$~mas and
$\Delta\delta=+21.4$~mas. In the case of LS~5039, due to its proper
motion (see \citealt{ribo02}), the source was found to be
$\Delta\alpha=+7.6$~mas and $\Delta\delta=-16.2$~mas away from the correlated
phase center, located at $\alpha_{\rm J2000.0}=18^{\rm h} 26^{\rm m}
15\fs05600$ and $\delta_{\rm J2000.0}=-14\degr 50\arcmin 54\farcs2400$ (from
\citealt{marti98}).

Regarding the VLA observations, the source 3C~286 was used for flux
density calibration, and the source J1825$-$1718 for phase calibration
and to phase up the array.

\section{Data reduction} \label{reduction}

We performed the post-correlation data reduction using the Astronomical Image
Processing System ({\sc aips}) software package, developed and maintained by
NRAO. The phased VLA position had to be corrected by 43~cm, according to later
geodetic measurements. The positions of J1825$-$1718 and
LS~5039 were corrected using the task {\sc clcor}. As recommended for
phase-referencing experiments, we applied ionospheric and Earth Orientation
Parameters corrections to the visibility data using the task {\sc clcor}. A
priori visibility amplitude calibration was done using the antenna gains and
the system temperatures measured at each station. We then used the fringe
finder to calibrate the instrumental phase and delay offsets. The fringe
fitting ({\sc fring}) of the residual delays and fringe rates was performed for
all the radio sources. We found very good solutions for 3C~345,
J1911$-$2006 and J1825$-$1718. Fringes for 15\% and 25\% of
the baselines were missing for LS~5039 and J1837$-$1532,
respectively. Typical data inspection and flagging were performed. The obtained
self-calibrated images of J1911$-$2006, which show a one-sided jet,
are very similar to those present in the VLBA calibrator source catalog,
assessing the reliability of the observations and the initial data reduction.

For each run, we performed standard phase and amplitude calibration and phase
self-calibration steps on LS~5039. As a compromise between angular
resolution and sensitivity, we  used a weighting scheme with robust 0 to
produce the images with the task {\sc imagr}. The obtained image had a
synthesized beam of $\sim$6.4$\times$2.2~mas in position angle (PA) of
$\sim0\degr$. Extended emission was visible for both runs. To enhance the
small-scale morphology, we produced final images using a convolving beam of
3.4$\times$1.2~mas in PA of $0\degr$ (similar to what would be obtained by
using robust $-$2 within {\sc aips} or a uniform weighting scheme within {\sc
difmap}). This convolving beam is the one we used previously in
\cite{paredes00}, allowing for a straightforward comparison of the images.

To measure the position of LS~5039 in both runs we used the
phase-referencing technique, and transferred the phases of the calibrator
J1825$-$1718, fitted with the task {\sc fring}, to our target source.
The peak flux density of this extragalactic phase-reference calibrator is
118~mJy~beam$^{-1}$ at 5~GHz, enough to provide useful phase information.
However, it suffers scattering and becomes resolved with the VLBA, providing
relatively noisy phase information on the longest baselines. We inspected and
screened the phase delays and delay rates of J1825$-$1718.
Tropospheric corrections were unsuccessful because the observation strategy was
not adequate to account for them. Phase-referenced images of the target source
with a synthesized beam of $\sim$4.9$\times$1.5~mas in PA $\sim0\degr$ were
obtained, and the peak position was measured with the {\sc aips} task {\sc
jmfit}.

\section{Results} \label{results}

\subsection{VLA} \label{VLA}

The VLA data of LS~5039 were compatible with a point-like source for
the obtained synthesized beam of $5.5\times3.7\arcsec$ in PA of $\sim-$4\degr.
We measured the flux density of the source every 30 minutes and obtained a mean
of 29.4~mJy with a standard deviation of $\sigma$=1.1~mJy during the 8 hours of
run~A, and 28.4~mJy with $\sigma$=0.7~mJy for run~B. Therefore,
LS~5039 displayed flux density variability below $4\%$ at the
1$\sigma$ level within each 8-hour run. Similar results were obtained within
4-hour halves. This guarantees that we can produce reliable VLBA images for
each of the entire observation periods, as well as for 4-hour intervals,
provided the morphology does not change within such intervals.

\subsection{VLBA} \label{vlba}

\begin{figure*}[t!] 
\center
\resizebox{1.0\hsize}{!}{\includegraphics[angle=0]{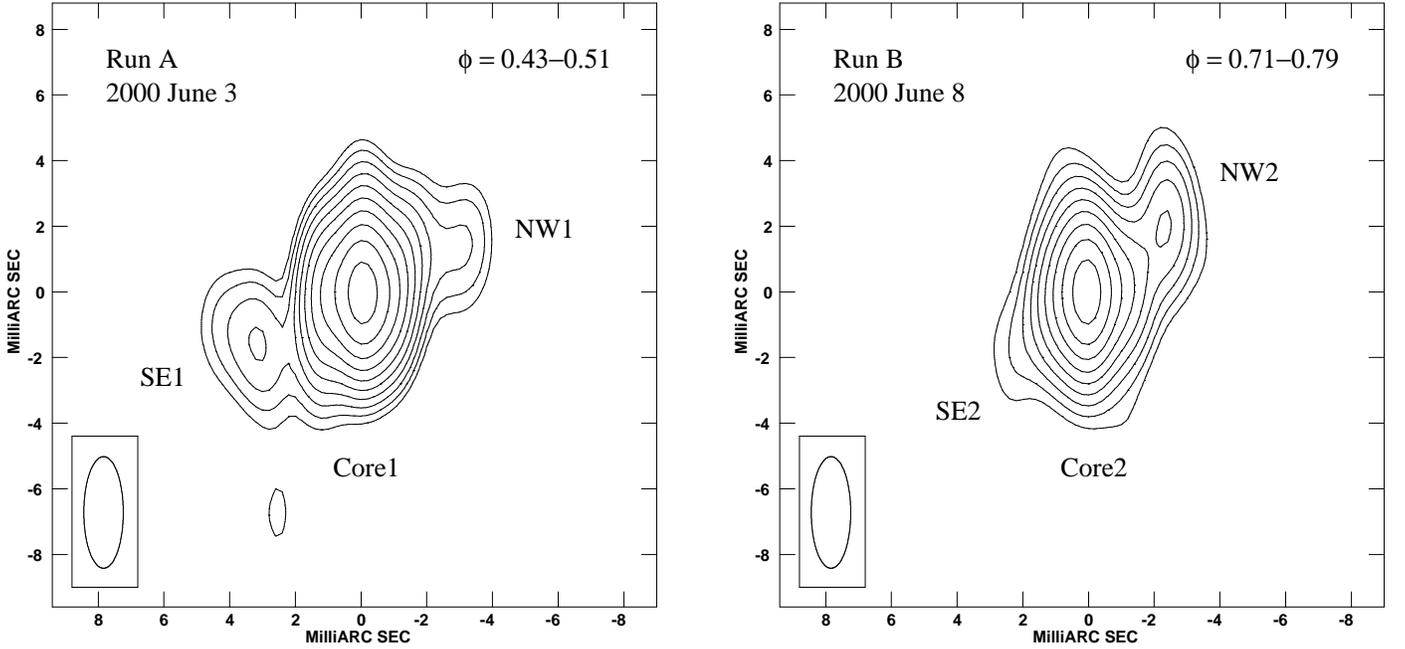}}
\caption{VLBA+phased VLA self-calibrated images of LS~5039 at 5~GHz
obtained on 2000 June 3 (left) and 8 (right). North is up and east is to the
left. Axes units are mas, and the (0,0) position corresponds to the source peak
in each image. The convolving beam, plotted in the lower left corner, has a
size of 3.4$\times$1.2~mas in PA of $0\degr$. The first contour corresponds to
5 times the rms noise of the image (0.08 and 0.11~mJy~beam$^{-1}$ for run~A and
B, respectively), while consecutive ones scale with $2^{1/2}$. The dates and
orbital phases are quoted in the images. There is extended radio emission that
appears nearly symmetric for run~A and clearly asymmetric for run~B, with a
small change of $\sim$10\degr\ in its position angle. We have labeled the
components fitted to the data, the parameters of which are listed in
Table~\ref{table:parameters}.}
\label{fig:vlba}
\end{figure*}

We show the final VLBA+phased VLA self-calibrated images in
Fig.~\ref{fig:vlba}. The image obtained for run~A displays a central core and
bipolar and nearly symmetric extended emission with PA $\simeq116\pm2\degr$,
with enhanced emission toward the southeast. The total flux density recovered
by the VLBA, obtained with the task {\sc tvstat} within {\sc aips}, is
$\sim$25~mJy, representing $\sim$85\% of the VLA value. The peak flux density
of the core is 10.5~mJy~beam$^{-1}$. The image is similar to the one obtained
with the same array in 1999 May 8, corresponding to orbital phases 0.12--0.15,
which showed a slightly more asymmetric extended emission in PA$\sim$125\degr,
as can be seen in \cite{paredes00}. In contrast, the image obtained for run~B
displays a core and bipolar but clearly asymmetric structure, with PA
$\simeq128\pm2\degr$ and enhanced emission toward the northwest. The total flux
density recovered is $\sim$24~mJy, or $\sim$85\% of the VLA value. The peak
flux density of the core is 10.5~mJy~beam$^{-1}$, as in run~A.

\begin{table*} 
\begin{center}
\caption{Parameters of the Gaussian components fitted to the data for each
8-hour run. Columns~3 and 4 list the peak and integrated flux densities of each
component. Columns~5 to 8 list the polar and Cartesian coordinates of the
components with respect to the peak position. Columns~9 to 11 list the sizes of
the Gaussian major and minor axes and the PA of the major axis. The parameters
PA and PA$_{\rm Axis}$ are positive from north to east.}
\label{table:parameters} 
\begin{tabular}{l@{~~~}l@{~~}r@{ $\pm$ }l@{}r@{ $\pm$ }l@{~~~~}r@{ $\pm$ }l@{~~~~}r@{ $\pm$ }l@{~~~~}r@{ $\pm$ }l@{~~~~}r@{ $\pm$ }l@{~~~~}r@{ $\pm$ }l@{~~~~}r@{ $\pm$ }l@{~~~~}r@{ $\pm$ }l@{}}
\hline
\hline
Run & Comp. & \multicolumn{2}{c}{Peak $S_{\rm 5~GHz}$} & \multicolumn{2}{c}{$S_{\rm 5~GHz}$} & \multicolumn{2}{c}{$r$}   & \multicolumn{2}{c}{PA}         & \multicolumn{2}{c}{$\Delta\alpha$} & \multicolumn{2}{c}{$\Delta\delta$} & \multicolumn{2}{c}{Maj. Axis} &\multicolumn{2}{c}{Min. Axis} & \multicolumn{2}{c}{PA$_{\rm Axis}$} \\
    &       & \multicolumn{2}{c}{[mJy~beam$^{-1}$]}    & \multicolumn{2}{c}{[mJy]}           & \multicolumn{2}{c}{[mas]} & \multicolumn{2}{c}{[$\degr$]} & \multicolumn{2}{c}{[mas]}          & \multicolumn{2}{c}{[mas]}          & \multicolumn{2}{c}{[mas]}     &\multicolumn{2}{c}{[mas]}     & \multicolumn{2}{c}{[$\degr$]} \\
\hline
A   & Core1 & 10.54 & 0.08 & 20.0 & 0.2 & \multicolumn{2}{c}{---} & \multicolumn{2}{c}{---} & \multicolumn{2}{c}{---} & \multicolumn{2}{c}{---} & 3.69 & 0.03 & 2.09 & 0.02 & 103 & 1 \\
    & SE1   &  1.11 & 0.08 &  2.6 & 0.2 & 3.67  & 0.08 & 115.9   & 1.7 &    3.30 & 0.07 &  $-$1.60 & 0.12 & 4.1 & 0.3 & 2.36 & 0.16 & 17 & 5 \\
    & NW1   &  0.88 & 0.08 &  1.5 & 0.2 & 3.29  & 0.09 & $-$63   & 2   & $-$2.92 & 0.08 &     1.52 & 0.14 & 3.6 & 0.3 & 1.99 & 0.18 &  3 & 6 \\
\hline
B   & Core2 & 10.45 & 0.11 & 17.6 & 0.3 & \multicolumn{2}{c}{---} &\multicolumn{2}{c}{---} &\multicolumn{2}{c}{---} &\multicolumn{2}{c}{---} &3.71 & 0.04 &1.8  & 0.2 &180 & 1 \\
    & SE2   &  0.75 & 0.11 &  1.8 & 0.4 & 2.8~~ & 0.2  &     129 & 5   &    2.17 & 0.13 & $-$1.8~~ & 0.3  & 4.6 & 0.7 & 2.1  & 0.3  &   1 & 7 \\
    & NW2   &  2.22 & 0.11 &  3.9 & 0.3 & 2.94  & 0.06 & $-$52.2 & 1.4 & $-$2.32 & 0.04 &     1.80 & 0.09 & 4.3 & 0.2 & 1.68 & 0.08 & 174 & 2 \\
\hline
\end{tabular}
\end{center}
\end{table*}

To characterize the extended emission we used {\sc uvfit} and {\sc jmfit}
within {\sc aips}, as well as model fitting tools within {\sc difmap} to check
the reliability of the obtained results. The preferred model for run~A data
consisted of three Gaussian components to account for the core (Core1), the
southeast (SE1) and the northwest (NW1) components. A similar model could fit
the data for run~B, although the southeast (SE2) component is marginally fitted
in this case. The fitted parameters are quoted in Table~\ref{table:parameters}.
As can be seen, the peak flux density of each component is $\sim$50\% of its
total flux density, although this is partially due to the convolving beam size
being smaller than the synthesized beam size. In particular, we stress that the
core components are not point-like.

Finally, we splitted the 8~hours of data in each run into 4-hour data sets. For
each of them, we performed a similar calibration process as the one described
above, and we modelled the final visibilities of the self-calibrated data. We
measured no significant morphological differences between the two halves in any
of the two runs. In particular, the distance between the fitted components
Core1 and SE1 is stable in 4 hours within the errors ($\sigma_{\alpha}^{\rm
A}=0.31$~mas, $\sigma_{\delta}^{\rm A}=0.62$~mas). This is also the case
between Core2 and NW2 components ($\sigma_{\alpha}^{\rm B}=0.55$~mas,
$\sigma_{\delta}^{\rm B}=0.25$~mas).

\subsection{Astrometry} \label{astrometry}

The phase-referenced image of LS~5039 obtained for run~A, not shown
here, displays a compact source with peak flux density of 6.7~mJy~beam$^{-1}$,
and a faint extended emission toward the southwest ($\sim$2~mJy~beam$^{-1}$;
the rms is 0.3~mJy~beam$^{-1}$). The measured position of LS~5039 for
run~A is $\alpha_{\rm J2000.0}=18^{\rm h} 26^{\rm m} 15\fs05653\pm0\fs00001$
(or $\pm$0.15~mas), and $\delta_{\rm J2000.0}=-14\degr 50\arcmin
54\farcs2564\pm0\farcs0015$ (or $\pm$1.5~mas) (see discussion below). We recall
that the reference coordinates are those of J1825$-$1718 listed in
Sect.~\ref{observations}, which have systematic uncertainties of $\pm3.0$~mas
and $\pm4.5$~mas, respectively. For checking purposes we splitted the data in
blocks of 4- and 1-hour lengths, made images for each block, and measured the
corresponding positions. The peak position of LS~5039 appears to move 
$\Delta\alpha=+0.1\pm0.1$~mas, $\Delta\delta=+2.8\pm0.2$~mas between the two
4-hour blocks, and slightly more in the 1-hour blocks. The observed direction
has a PA of $1.7\pm1.7\degr$, which is the same as the line joining the
positions of the source and the phase-reference calibrator
J1825$-$1718. On the other hand, the expected error for differential
astrometry is given by the separation $d$ in degrees from the phase-reference
source and the offset $\Delta$ of its correlated position, according to
$\Delta\times(d/180)\times\pi$ (see \citealt{walker99}). Plugging our offset of
60.9~mas and a distance of 2\fdg47 we obtain an error of 2.6~mas, very similar
to the observed displacement of LS~5039. A similar procedure using
4-hour blocks for the astrometric check source J1837$-$1532
($d=3\fdg38$) reveals motion of the peak of this extended source by
$3.2\pm0.4$~mas at PA=$46\pm8\degr$, as expected. Therefore, these secular
motions appear to be purely instrumental. The errors assigned to the
coordinates of LS~5039 quoted above are half of the total secular
motion measured in 1-hour blocks.

The phase-referenced image of run~B, not shown here, reveals a double source in
the eastwest direction. The two components have similar peak flux densities of
$3.6\pm0.1$~mJy~beam$^{-1}$, much lower than the one of the self-calibrated
core, and are separated $2.5\pm0.1$~mas in PA of $87\pm5\degr$. Moreover, 4-
and 1-hour blocks reveal a fading of the western component and a brightening of
the eastern one along the run (as well as a similar secular motion as in
run~A). This symmetric double structure is in contrast to the asymmetric
structure seen in the self-calibrated image shown in Fig.~\ref{fig:vlba}-right.
Tropospheric errors, which affect the phase-referenced image and cannot be
accounted for, can easily split Core2 into the observed double source.
Therefore, the precise position of the peak of LS~5039 cannot be
measured in run~B.

\section{Discussion} \label{discussion}

The observations of LS~5039 reported here, obtained with the VLBA on
two runs separated by 5 days, show a changing morphology at mas scales. In both
runs there is a core component with a constant flux density within errors, and
elongated emission with a PA that changes by $12\pm3\degr$ between both runs.
The source is nearly symmetric in run~A and asymmetric in run~B (see
Fig.~\ref{fig:vlba}).

In the microquasar scenario, and assuming ballistic motions of adiabatically
expanding plasma clouds without shocks (see \citealt{mirabel99}), the
morphology of run~A can be interpreted as a double-sided jet emanating from a
central core with the southeast component as the approaching one. The relative
distances of the components to the core would imply a bulk motion with velocity
$\beta>0.06\pm0.02$, whereas the flux asymmetry implies $\beta>0.08\pm0.02$ for
discrete ejections. The angle $\theta$ between the jet and the line of sight is
restricted to $<$87\degr. These results are similar to those previously found
with VLBA and EVN images \citep{paredes00,paredes02}. We can compute, for
different $\theta$ angles, the expected displacement of the SE1 component with
respect to Core1 in 4 hours using the measured value for the flux asymmetry of
$\beta\cos\theta=0.08$, and compare it to the upper limits for this
displacement quoted at the end of Sect.~\ref{vlba}. This provides a constraint
of $\theta<48\degr$ at the 3$\sigma$ level. In contrast, in run~B the northwest
component appears to be the approaching one emanating from the core. The
distances from Core2 to the components NW2 and SE2 are very similar and do not
imply any significant relativistic motion. The flux asymmetry implies
$\beta>0.11\pm0.03$ and $\theta<84\degr$ for discrete ejections. As for run~A,
our measured upper limits of any displacement in 4 hours imply $\theta<45\degr$
at the 3$\sigma$ level. In conclusion, the lack of proper motions implies that,
for the measured flux asymmetries, the SE1 jet should be pointing at
$\theta<48\degr$, and the NW2 jet at $\theta<45\degr$. In this context, jet
precession is needed to explain this behavior. If the precession axis is close
to the plane of the sky, as in SS~433 \citep{blundell04}, the precession cone
should have a semi-opening angle $>45\degr$ to fulfill the $\theta$ limits
quoted above. If the precession axis is close to the line of sight, a small
precession of a few degrees could explain the images of runs A and B. However,
in both cases the PA of the jet should vary considerably, in contrast to the
small range covered by all observed values at mas scale, between 115 and
140\degr\ \citep{paredes00,paredes02}.

Alternatively, the morphology detected in run~B could be the result of a
discrete ejection with core suppression where Core2 is the approaching
component and NW2 the receding one, while there is no radio emission at the
origin of the ejection. In this case, the flux asymmetry implies
$\beta>0.21\pm0.01$ and $\theta<78\pm1\degr$. The measured upper limits of the
relative motion of these components in 4 hours provide a strong constraint of
$\theta<14\degr$ at the 3$\sigma$ level. Imposing $\beta\cos\theta=0.21$, the
origin of the discrete ejection (i.e., the position of the binary system or the
position of Core1 in run~A) should be placed at $\Delta\alpha=-1.4$~mas and
$\Delta\delta=+1.1$~mas from Core2 in Fig~\ref{fig:vlba}. Although this would
be self-consistent if the precise position of Core2 corresponds to the east
component of the phase-referenced image of run~B, it could be a chance
coincidence produced by tropospheric errors. One should also invoke jet
precession on short timescales to explain the change in PA of the extended
emission. However, large X-ray and radio flux density variations are observed
in microquasars displaying discrete ejections (see \citealt{fender06} and
references therein), while the peak and total radio flux densities of
LS~5039 are strikingly constant (see also, \citealt{ribo99};
\citealt{clark01}), and there is no evidence of an X-ray flare in 11.5 years of
{\it RXTE}/ASM data.

In the young nonaccreting pulsar scenario, the different morphologies we have
detected at different orbital phases could be due to the change of the relative
positions between the pulsar and the companion star along the orbit (see
\citealt{dubus06} for details). However, observations at different orbital
phases have always revealed a very similar PA for the extended emission, which
can only be observed if the binary system is seen nearly edge on. On the other
hand, the absence of X-ray eclipses places an upper limit of $i\la$75\degr\ in
this scenario \citep{dubus06}. Therefore, these two restrictions imply an
inclination angle that should be close to the upper limit of 75\degr. We note
that this is much higher than the inclination angle if the binary system is
pseudo-synchronized \citep{casares05}. 

In conclusion, a simple and shockless microquasar scenario cannot easily
explain the observed changes in morphology. On the other hand, an
interpretation within the young nonaccreting pulsar scenario requires the
inclination of the binary system to be very close to the upper limit imposed by
the absence of X-ray eclipses. Precise phase-referenced VLBI observations
covering a whole orbital cycle are necessary to trace possible periodic
displacements of the peak position, expected in this last scenario, and to
obtain morphological information along the orbit. These will ultimately reveal
the nature of the powering source in this gamma-ray binary.

\begin{acknowledgements}

We thank an anonymous referee for useful comments that helped to improve this paper.
We are grateful to Simon Garrington for allowing us to schedule dedicated 
MERLIN observations of possible phase-reference sources. We are grateful to
Craig Walker for his advice in the use of possible calibrators. The position of
J1825$-$1718 was provided by observations from the joint NASA/USNO/NRAO
geodetic/astrometric observing program. We thank Bob Campbell, Eduardo Ros and
Jos\'e Carlos Guirado for useful discussions about the phase-referencing
problems.
The NRAO is a facility of the National Science Foundation operated under
cooperative agreement by Associated Universities, Inc.
Part of the data reduction was done at JIVE with the support of the European
Community - Access to Research Infrastructure action of the Improving Human
Potential Programme, under contract No. HPRI-CT-1999-00045.
This research has made use of the NASA's Astrophysics Data System Abstract 
Service, and of the SIMBAD database, operated at CDS, Strasbourg, France.
M.R., J.M.P., and J.M. acknowledge support by DGI of the Spanish 
Ministerio de Educaci\'on y Ciencia (MEC) under grants 
AYA2007-68034-C03-01 and AYA2007-68034-C03-02 and FEDER funds.
M.R. acknowledges financial support from MEC through a \emph{Ram\'on y Cajal}
fellowship.
J.M. is also supported by Plan Andaluz de Investigaci\'on of Junta de 
Andaluc\'{\i}a as research group FQM322.

\end{acknowledgements}

\end{document}